\documentclass[aps,prb,twocolumn,groupedaddress,showpacs,floatfix,altaffilletter]{revtex4-1}
\usepackage{graphicx}
\usepackage{grffile}
\usepackage{amsmath}
\usepackage{amssymb}
\usepackage{bm}
\usepackage{color}
\usepackage[dvipsnames]{xcolor}
\usepackage{amsmath,amssymb,amsfonts}
\usepackage{epsfig}
\usepackage{times}
\usepackage{gensymb}
\usepackage{float}
\usepackage[colorlinks,bookmarks=false,citecolor=blue,linkcolor=red,urlcolor=blue]{hyperref}

\newcommand{\fref}[1]{Fig.~\ref{#1}}
\newcommand{\eref}[1]{Eq.~(\ref{#1})}
\newcommand{\sref}[1]{Sec.~(\ref{#1})}
\newcommand{\tref}[1]{Table~\ref{#1}}

\begin{document}

\title{Electronic and Magnetic Properties of the Candidate Magnetocaloric-Material Double Perovskites La$_2$MnCoO$_6$, La$_2$MnNiO$_6$ and La$_2$MnFeO$_6$}

\author{C.~Gauvin-Ndiaye$^{1}$, T. ~E.~Baker$^{1}$, P.~Karan$^{1}$, {\'E}.~Mass{\'e}$^1$, M. Balli$^{1}$, N. Brahiti$^{1}$, M. A. Eskandari$^{1}$, P. Fournier$^{1,2}$, A.-M.S.~Tremblay$^{1,2}$, and R.~Nourafkan$^{1}$}
\affiliation{$^1$Institut quantique, D{\'e}partement de Physique and RQMP, Universit{\'e} de Sherbrooke, Sherbrooke, Qu{\'e}bec, Canada  J1K 2R1}
\affiliation{$^2$Canadian Institute for Advanced Research, Toronto, Ontario, Canada M5G 1Z8}
\date{\today}
\begin{abstract}
The search for room-temperature magnetocaloric materials for refrigeration has led to investigations of double perovskites. In particular, a puzzle has appeared in the La$_2$MnNiO$_6$, La$_2$MnCoO$_6$ and La$_2$MnFeO$_6$ family of compounds. They share the same crystal structure, but while La$_2$MnNiO$_6$ and La$_2$MnCoO$_6$ are ferromagnets below room temperature, La$_2$MnFeO$_6$, contrary to simple expectations, is a ferrimagnet. To solve this puzzle, we use density-functional theory calculations to investigate the electronic structure and magnetic exchange interactions of the ordered double perovskites. Our study reveals the critical role played by local electron-electron interaction in the Fe-$d$ orbital to promote the Fe$^{3+}$ valence state with half-filled $d$-shell over Fe$^{2+}$ and to establish a ferrimagnetic ground state for La$_2$MnFeO$_6$. The importance of Hund's coupling and Jahn-Teller distortion on the Mn$^{4+}$ ion is also pointed out. Exchange constants are extracted by comparing different magnetically ordered states. Mean-field and classical Monte-Carlo calculations on the resulting model give trends in $T_C$ that are in agreement with experiments on this family of materials.

\end{abstract}

\maketitle
\section{Introduction}
The magnetocaloric effect leads to an increase of the temperature when certain materials are exposed to a magnetic field and decreases when they are removed from it. In order to be suitable for room temperature magnetic refrigeration, magnetocaloric materials need to exhibit a large change in magnetization around room temperature. The most interesting materials for this technology are hence ferromagnets with a high total moment per formula unit that undergo a magnetic phase transition to a paramagnetic state at room temperature.\cite{ref-mce} Recently, double perovskite have been given considerable attention for this technology because of  the low-production cost, stronger spin-phonon interactions, higher chemical stability, and better insulating properties.

Double perovskites La$_2$MnNiO$_6$ (LMNO) and La$_2$MnCoO$_6$ (LMCO) exhibit near room temperature ferromagnetism, with a Curie temperature of $T_c\simeq 280$~K~\cite{ADMA:ADMA200500737} and $T_c\simeq 226$~K~\cite{PhysRevB.67.014401}, respectively. From a practical point of view,  LMNO has  a refrigerant capacity  similar to that of gadolinium~\cite{doi:10.1063/1.4874943} and would be a promising candidate for magnetocaloric refrigeration if one could increase its  Curie temperature through  chemical substitution or thermal treatment. The low-temperature valence states of Mn and Ni ions in La$_2$MnNiO$_6$ are tetravalent (Mn$^{4+}$) and divalent (Ni$^{2+}$), with magnetic moments of $3\mu_B$ and $2\mu_B$ respectively, for a total of $5\mu_B$ per formula unit (f.u.).~\cite{doi:10.1063/1.2226997} LMCO is also a ferromagnet with tetravalent Mn and divalent Co and a total moment of 6$\mu_B$/f.u.

This motivates the study of La$_2$MnFeO$_6$ (LMFO): naively, one may think that substituting Ni or Co with Fe may increase $T_C$ since Fe has a magnetic moment of $5\mu_B$  in its trivalent high spin configuration, i.e, Fe$^{3+}$ with half-filled $d$ shells. Ferromagnetic LMFO with trivalent Mn and Fe would then have a higher total moment (9$\mu_B$/f.u.) and possibly a higher Curie temperature than both LMCO and LMNO.
However, experimental results show that LMFO is a ferrimagnet with anti-parallel moments on neighboring Mn and Fe sites.~\cite{PhysRevB.91.054421, PhysRevB.60.R12561, doi:10.1063/1.1343844} 


The difference in the ground state magnetic orders in LMFO and LMNO/LMCO is puzzling because all of these materials share the same crystal structure. The resulting double perovskite A$_2$B'B''O$_6$ structure stems from the perovskite structure ABO$_3$  where half of the transition metal sites (B) are occupied by Mn ions and the other half by Ni, Fe or Co ions.~\cite{VASALA20151}  In such a structure resulting from the solid solution of AB'O$_3$ and AB''O$_3$, the interaction between neighboring moments occurs through the oxygen-mediated superexchange interaction. Superexchange facilitates hopping of the electrons from the oxygen $p$ shells to the magnetic ions $d$ shell, leading to a reduction of the total energy due to a kinetic energy advantage.  It can be antiferromagnetic (AFM) or ferromagnetic (FM), depending on structural parameters such as the B'-O-B'' bond angle and the B'-O-B'' bond length. Furthermore, crystal field splitting, Hund's coupling and on-site electron-electron repulsion can also influence the superexchange interaction. The roles played by the latter parameters depend on the $d$ shell occupancy. 

\begin{figure}
\includegraphics[width=0.8\columnwidth]{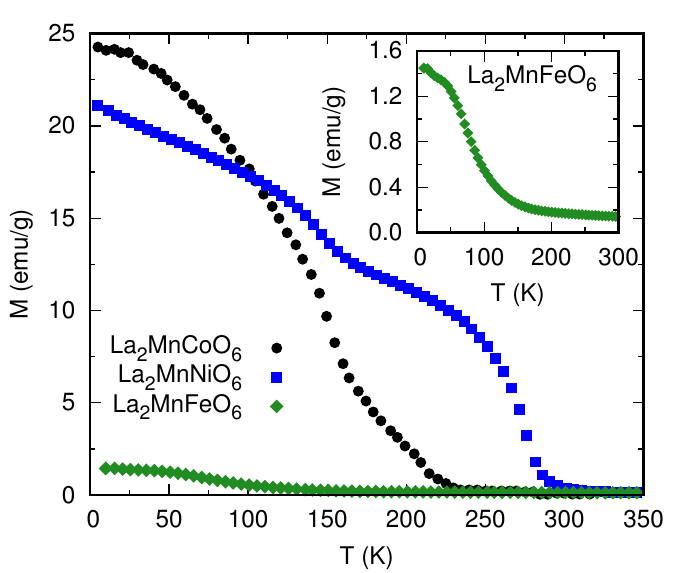}\\
\caption{(Color online) Magnetization as a function of temperature in bulk La$_2$MnCoO$_6$ (black circles); La$_2$MnNiO$_6$ (blue squares) and La$_2$MnFeO$_6$ (green triangles). Inset: zoom on the magnetic transition in LMFO.}
    \label{fig:MvsT}
\end{figure}

Finally, the properties of double perovskites are strongly influenced by the level of cationic order. In the ordered phase, the 3d metal cations crystallize in the so-called rock-salt structure in which Mn and Ni (Fe, Co) atoms occupy alternate positions in each spatial direction. In the pristine double perovskites, the space group symmetry becomes $P2_1/c$.~\cite{doi:10.1063/1.4874943, doi:10.1063/1.4893721, PhysRevB.67.014401} In the disordered phase, for which not only B'-O-B'' but also B'-O-B' and B''-O-B'' bonds are present, the materials have been reported to crystallize in the $Pbnm$ space group.~\cite{C5RA24092A,doi:10.1063/1.4874943, PhysRevB.67.014401} For example in La$_2$MnCoO$_6$, the ferromagnetic Mn-O-Co bonds are diluted in a matrix of antiferromagnetic Co-O-Co and Mn-O-Mn bonds. Depending on the growth parameters, we measure two distinct transitions in a bulk sample as shown in ~\fref{fig:MvsT} for La$_2$MnCoO$_6$ and La$_2$MnNiO$_6$, demonstrating a mixture of the ordered and the disordered domains. Tuning the growth parameters allows us to tune the proportion of both phases and eventually to remove completely the low temperature transition associated to the cation-disordered phase and leaving only the high temperature transition of the cation-ordered phase.\cite{Singh09d}
In the present theoretical study, we will focus on the magnetic properties of the cation-ordered phase while trying to explain the absence of a high temperature transition in La$_2$MnFeO$_6$ as shown also in \fref{fig:MvsT}.

%
%

The understanding of the microscopic mechanisms at play in LMFO could lead to the design of new materials that could be more suitable for magnetic refrigeration than LMNO and LMCO.

We first investigate the ground states of LMNO, LMCO and LMFO with density functional theory (DFT) calculations in Sec.~(\ref{Sec:Results}). We compare the structural, electronic and magnetic properties of the three materials in order to understand the impact of electron-electron interaction and structural distortion on the ground state of LMFO. In Sec.~(\ref{section:exchange}), electronic structure calculations for different types of magnetically ordered ground states allow us to extract exchange coupling constants and corresponding mean-field transition temperatures. In Sec.~(\ref{section:CMC}), we improve the estimates of the Curie temperature using Monte-Carlo calculations. The method specific to each section is described in the opening subsection. 
\section{Electronic structure for L\lowercase{a}$_2$M\lowercase{n}N\lowercase{i}O$_6$, L\lowercase{a}$_2$M\lowercase{n}C\lowercase{o}O$_6$ and L\lowercase{a}$_2$M\lowercase{n}F\lowercase{e}O$_6$} \label{Sec:Results}
We first describe the method, then present results for LMNO and LMCO in Sec.~(\ref{section:LMNO_LMCO}). The puzzling case of LMFO is presented in Sec.~(\ref{section:LMFO}).
\subsection{Method}\label{:Sec.Method}
The  GGA(+U) calculations are performed within the full-potential all electron basis set as implemented in the WIEN2k code, using the PBE functional.~\cite{wien2k} In GGA+U calculation, the effective interaction $U_{eff}=U-J$ has been set to $3.0$~eV for Mn, Ni, Co and Fe $d$ orbitals (except when specified otherwise in the text). We use the GGA+U(SIC) method with an approximate correction for the double-counting.~\cite{PhysRevB.48.16929} Structure optimization is performed on LMNO, LMCO and  LMFO using the $P2_1/c$ space group. In order to confirm the ground state magnetic order, structure optimization is performed using two different magnetic alignments of Mn and Fe moments: ferromagnetic (FM), and (G-type) antiferromagnetic (AFM). A plane-wave cut-off of R\textsubscript{mt}$\cdot$K\textsubscript{max}$=8.1-8.3$ and a ${\bf k}$-mesh of $50$ points in the Brillouin zone are used for self-consistent calculations presented in \sref{section:LMNO_LMCO} and \sref{section:LMFO}. We check the convergence with respect to the number of ${\bf k}$ up to $250$ points in the irreducible Brillouin zone and to R\textsubscript{mt}$\cdot$K\textsubscript{max} cutoff up to $8.1$ for LMNO and LMCO, and $8.3$ for LMFO. 
In supercell calculations, a R\textsubscript{mt}$\cdot$K\textsubscript{max}=7 and 60 k-points were used (\sref{section:exchange}).

\subsection{L\lowercase{a}$_2$M\lowercase{n}N\lowercase{i}O$_6$ and L\lowercase{a}$_2$M\lowercase{n}C\lowercase{o}O$_6$}
\label{section:LMNO_LMCO}
\begin{figure*}
    \centering
    \begin{tabular}{ccc}
         \includegraphics[width=0.33\textwidth]{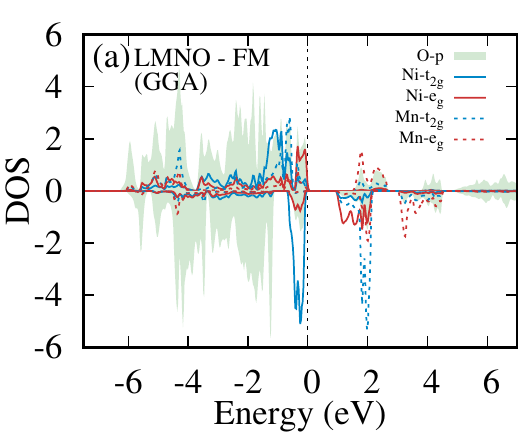} & 
         \includegraphics[width=0.33\textwidth]{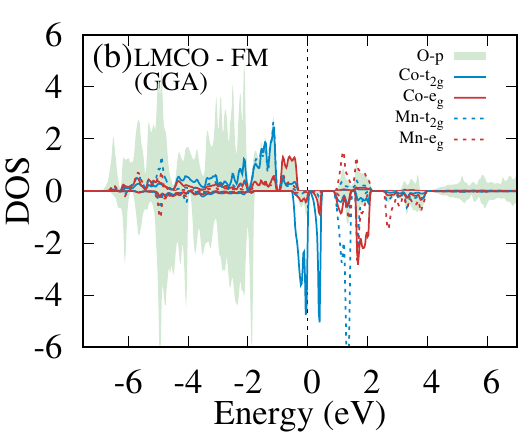} & 
         \includegraphics[width=0.33\textwidth]{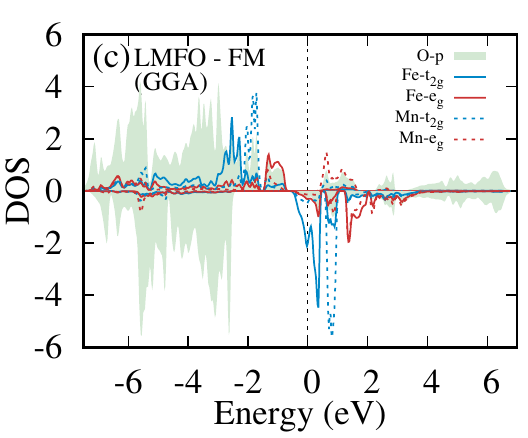} \\
         \includegraphics[width=0.33\textwidth]{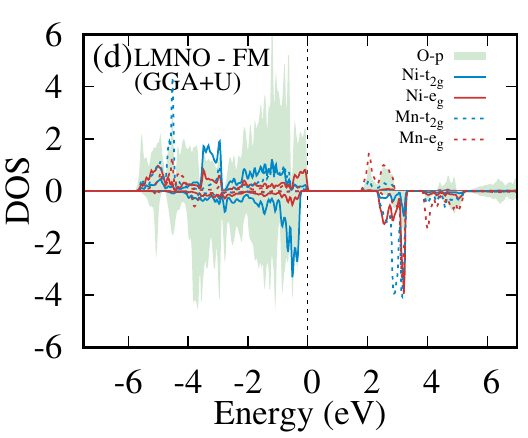} & 
         \includegraphics[width=0.33\textwidth]{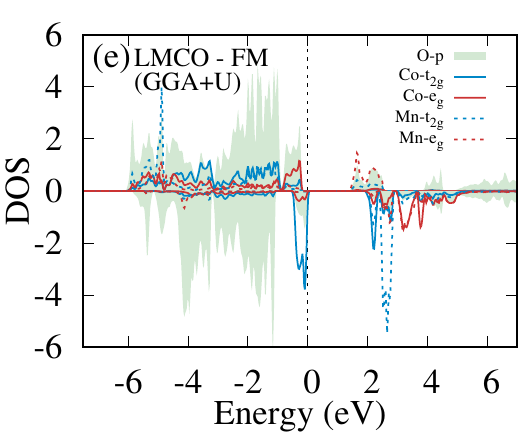} & 
         \includegraphics[width=0.33\textwidth]{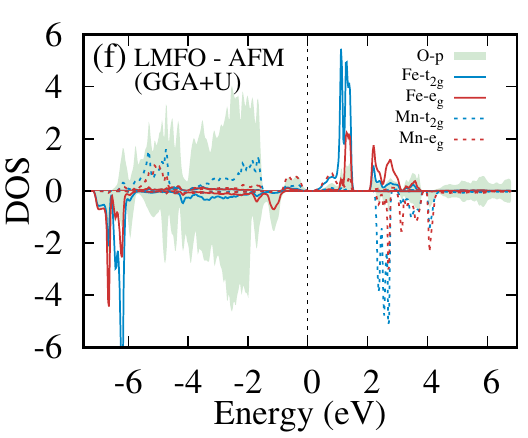} \\\\
         \includegraphics[width=0.33\textwidth]{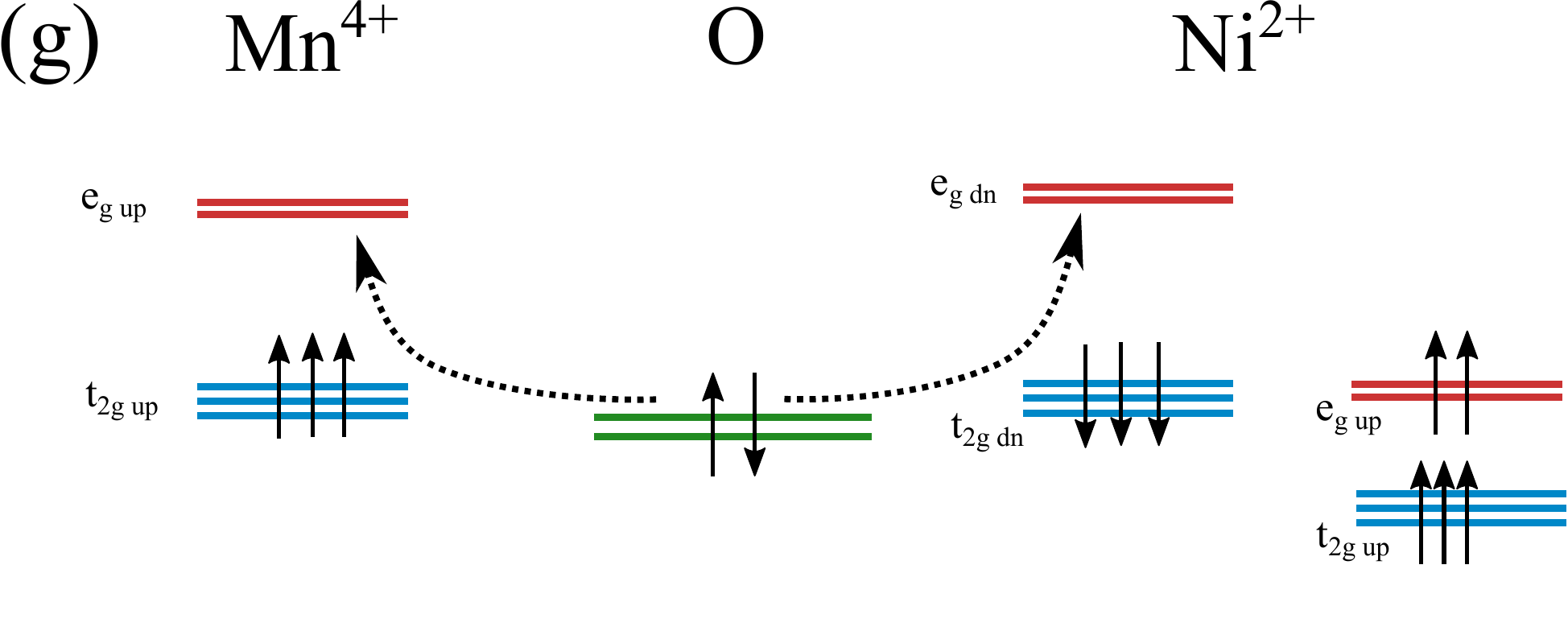} & 
         \includegraphics[width=0.33\textwidth]{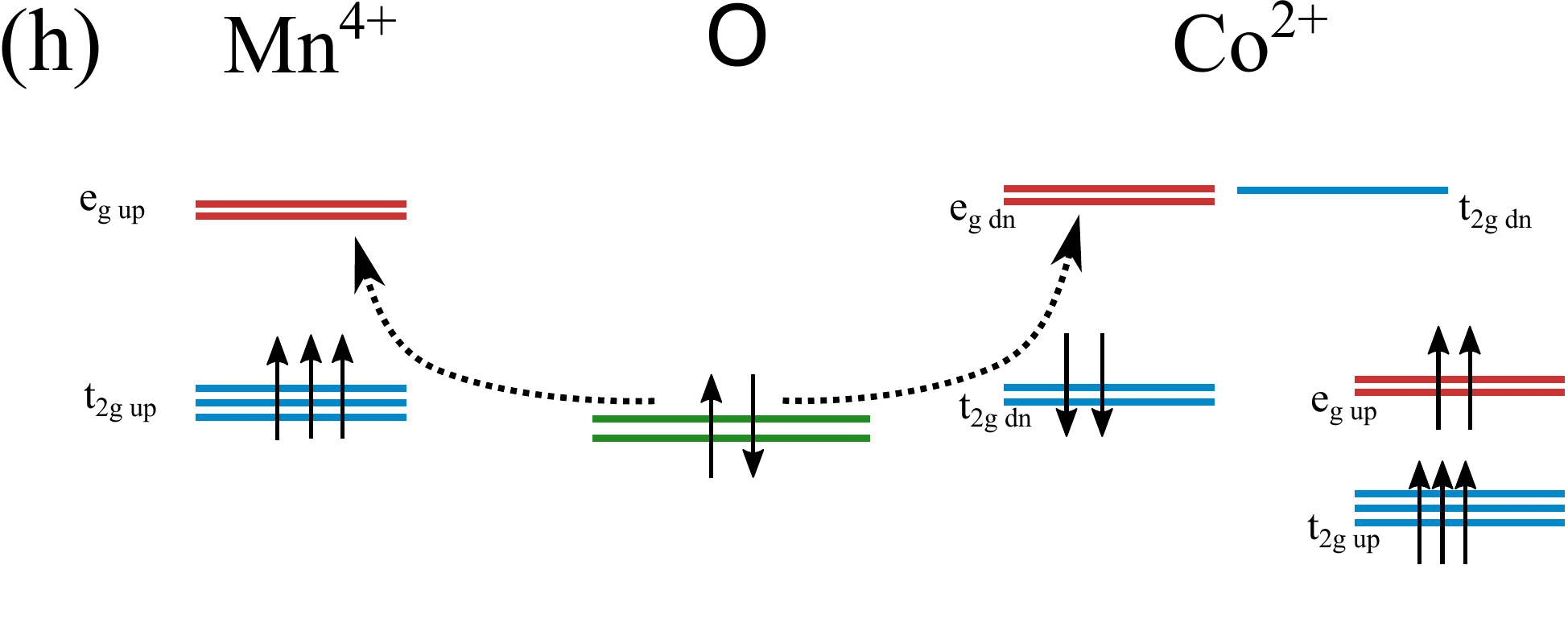} &
         \includegraphics[width=0.33\textwidth]{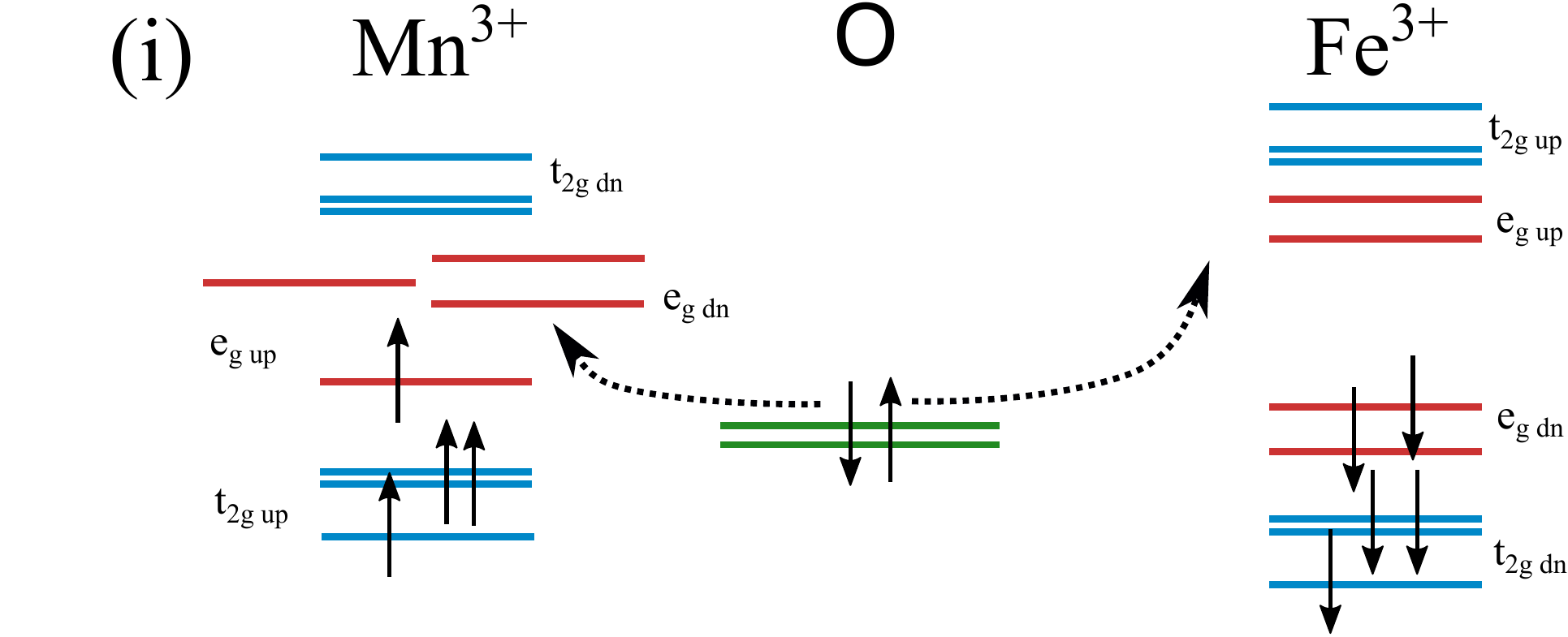} \\
    \end{tabular}
    \caption{(Color online) Spin-resolved partial density of states for Mn-$e_g$, Mn-$t_2g$, Ni (Co, Fe)-$e_g$, Ni (Co, Fe)-$t_{2g}$ and O-$p$ from GGA calculations (top row) and GGA+U (middle row) calculations. GGA+U calculations are performed with $U_{eff}=3$eV. The upper part in each panel is majority-spin DOS result, and the lower the minority-spin
one. GGA calculations predict an insulating ferromagnetic ground state for (a) LMNO and (b) LMCO, and a metallic ferromagnetic ground state for (c) LMFO.
 GGA+U calculations predict an insulating ferromagnetic ground state for (d) LMNO and (e) LMCO, and an insulating ferrimagnetic ground state for (f) LMFO.
    Bottom row: Schematic representation of the superexchange interaction in (g) LMNO, (h) LMCO and (i) LMFO. The figures represent schematically the weight of each orbital with respect to the others and are derived from the partial DOS plots and partial charge data. In LMFO, $e_g$ electrons are more itinerant than the $t_{2g}$ electrons and therefore repulsion $U$ has a smaller impact on them. This leads to a smaller energy splitting between up- and down-spin states of $e_g$ electrons and in turn to positioning of $e_g$ states between $t_{2g}$ states.
    }
    \label{fig:exchange_and_dos}
\end{figure*}

Starting with La$_2$MnNiO$_6$, our ab initio electronic structure calculations confirm that the orbital occupancy of Mn-$d$ and Ni-$d$ are very close to the nominal ones. The oxidation state of Mn and Ni are Mn$^{4+}$ and Ni$^{2+}$ with $3d^3_{\sigma}d^0_{\bar{\sigma}}$ and $3d^5_{\sigma}d^3_{\bar{\sigma}}$ electronic configurations, where $\bar{\sigma}=-\sigma$. Hence, the predicted total magnetic moment is $\simeq 5 \mu_B$/f.u., which is in good agreement with experimental data.~\cite{doi:10.1063/1.4874943} The energy difference between AFM and FM configurations is $0.13$ eV/f.u. in GGA calculations, and $0.16$ eV/f.u. in GGA+U calculations. \fref{fig:exchange_and_dos} (a) and (d) show partial density of states (DOS) of LMNO calculated from GGA and GGA+U methods respectively. The calculation is done in a FM magnetic moment configuration. Partial DOS of up (down)-component is denoted by positive (negative) value. Both methods predict an insulating ground state. 
Although the charge gap increases between GGA and GGA+U calculations, the partial charge occupations and the magnetic moments of the transition metal ions do not depend sensitively on effective Coulomb interaction. The  overlap between partial DOS of transition metal and oxygen above the Fermi level indicates which virtual hopping processes can be realized in the system.    
The partial DOS above the Fermi level illustrates good overlap between Ni-$e_g$ and O-$p$, in particular for down spin. Similarly, Mn-$e_g$ up and O-$p$ partial DOSs overlap well above the Fermi level.  
The optical gap predicted by GGA calculations is around $0.8$~eV and, above the Fermi level, the states are dominated by the Mn-$e_g$ orbitals, as shown in \fref{fig:exchange_and_dos} (a). 
Including the electron-electron interaction through GGA+U calculation pushes the Mn-$e_g$ states to higher energies, hence, just above the Fermi level, the states become dominated by the O-$p$ orbitals. The charge gap also increases to $1.7$~eV, as shown in \fref{fig:exchange_and_dos} (d). 

In LMNO, the main factor determining the ground state magnetic structure is the kinetic energy gain due to superexchange interaction. This is illustrated schematically in \fref{fig:exchange_and_dos}~(g) where the position of the atomic levels roughly corresponds to their position in the DOS.  Assuming that the electronic configuration of Mn is $3d^3_{\uparrow}d^0_{\downarrow}$, then the O-$p_{\uparrow}$ electron has a higher hopping amplitude to Mn-$e_g^{\uparrow}$ states than the O-$p_{\downarrow}$ electron because Hund's coupling between $t_{2g}$ and $e_g$ states of Mn favors parallel alignment. On the other hand, the $t_{2g}$ states of Ni are fully occupied. The extra two electrons reside in $e_g$ states and have parallel spins due to Hund's coupling. Since O-$p_{\downarrow}$ contributes in a superexchange mechanism with the $e_g$ states of Ni, a spin-up configuration is preferred for Ni-$e_g$ electrons. This explains why La$_2$MnNiO$_6$ is a ferromagnet. It is worth noting that the alternative Mn$^{3+}$ and Ni$^{3+}$ oxidation states would have led to a AFM configuration, but it is not energetically favorable due to the large crystal field splitting on Mn ions (see \sref{section:exchange}). 

GGA and GGA+U calculations performed on LMCO lead to similar results. \fref{fig:exchange_and_dos} (b) and (e) show the partial DOS of LMCO. The charge gap is very small in GGA, around $0.1$~eV. Adding  $U$ splits the Co-$t_{2g}$ states and  enhances the charge gap to $\simeq 1.3$~eV.
Both types of calculations predict a ferromagnetic insulating ground state with $\simeq6\mu_B$/f.u., which is in agreement with the experimental data.~\cite{doi:10.1063/1.4893721} The energy difference between AFM and FM configurations is $0.14$ eV/f.u. in GGA calculations, and $0.18$ eV/f.u. in GGA+U calculations. The superexchange mechanism in LMCO is similar to the one described for LMNO, with the difference that Co has a 3$d^5_\uparrow d^2_\downarrow$ electronic configuration (see \fref{fig:exchange_and_dos} (h)). Similarly to LMNO, properties such as charge occupation or partial moments of Mn and Co atoms do not vary sensibly between GGA and GGA+U calculations. 

By investigating LMNO and LMCO, we identify a key player in promoting the FM ground state:  it is the Mn$^{4+}$ oxidation with three electrons in $t_{2g}$ states that allows the Hund's coupling to become effective and to reduce the total energy of FM state with respect to AFM state. 

\subsection{L\lowercase{a}$_2$M\lowercase{n}F\lowercase{e}O$_6$}
\label{section:LMFO}
In light of the previous results, we move to the interesting case of LMFO. The GGA calculations predict a metallic ferromagnetic ground state for LMFO with finite Fe-$d$ spectral weight at the Fermi level as seen from the density of states of \fref{fig:exchange_and_dos} (c). The total moment predicted by GGA calculations is $\sim7\mu_B$/f.u., which corresponds to high-spin Mn$^{4+}$ and Fe$^{2+}$ states. The energy difference between the AFM and the FM configuration is $0.09$ eV/f.u. These results are not in agreement with the available experimental data. Indeed, experiments on high-order films (B-site order $\simeq63\%$) have found LMFO to be an insulating ferrimagnet with a total moment of $1.3\mu_B$/f.u. and a band gap of $1.1$ eV. ~\cite{PhysRevB.91.054421}
Adding electron-electron interaction in Fe-$d$ and Mn-$d$ orbitals in the GGA+U framework splits the spectral weight to lower and upper Hubbard band and opens a spectral gap as shown in \fref{fig:exchange_and_dos}~(f). Moreover, GGA+U calculations predict an AFM ground state with antiparallel alignment of the moments on Mn and Fe sites, and Mn$^{3+}$ and Fe$^{3+}$ states. The energy of the FM configuration is $0.11$ eV/f.u. higher than the energy of the AFM configuration. The relaxed structure shows Jahn-Teller distortion due to the presence of Mn$^{3+}$ ions. The predicted magnetic moment is $~1 \mu_B$/f.u., which is in agreement with $1.3\mu_B$  obtained in experiment for high-order films.~\cite{PhysRevB.91.054421} The band gap is around 0.2 eV, which is smaller than the experimental band gap 1.1 eV. However, as one can see from \fref{fig:exchange_and_dos} (f), the immediate spectral weights around the Fermi level are small, leading to a vanishingly small optical response for photon frequencies causing a transition between them. Therefore, the optical gap measured in experiment appears to be larger than what is calculated. Alternatively, the electron-electron interactions can be stronger than what we have considered here. By comparing calculated optical absorption edge with a measured one, we conclude that $U=3$ eV is a good approximation. 
The spectral weight above the Fermi level is predominantly given by Mn-$t_{2g}$, Fe-$e_g$ and Fe-$t_{2g}$ which is in agreement with the experimentally obtained spectra.~\cite{PhysRevB.91.054421} 
Experimentally, B-site order tends to appear in A$_2$B'B''O$_6$ double perovskites when the ions on B' and B'' sites have a charge difference larger than $2$ and an ionic radius difference larger than $0.17$\AA.\cite{ref-a2bbo6} The GGA+U predicted charges Mn$^{3+}$ and Fe$^{3+}$ could explain why no experiment has reported perfectly ordered LMFO. 

One can note that electron-electron interactions in the magnetic ions $d$ shells have a crucial impact on the ground state in LMFO, which is not the case for LMNO or LMCO.  
The oxidation state for Fe goes from $2^+$ ($d^6$) in GGA calculations to $3^+$ ($d^5$) in GGA+U calculations for the AFM phase. Hence, strong electron-electron interactions prevent double occupancy in Fe-$d$ shells. 
Moreover, using a larger value of $U_{eff}=6$ eV in both magnetic ions $d$ shells stabilizes the AFM ground state by further increasing the ground state total energy difference between FM and AFM phases.

A simple picture of the superexchange mechanism in LMFO can help understand its AFM ground state. Let us assume that Mn attains $4^+$ oxidation state in  La$_2$MnFeO$_6$, as is the case in LMNO. This oxidation state for Mn requires $2^+$ oxidation state for Fe with a $3d^5_{\sigma}d^1_{\bar{\sigma}}$ electronic configuration, where one of the Fe-$d$ orbital is doubly occupied. Such a double occupancy costs a large amount of energy for the system. Hence, the system avoids this energy cost by selecting  Mn$^{3+}$ and Fe$^{3+}$ oxidation states if the potential energy cost due to double occupancy of one Fe-$d$ orbital is larger than crystal field splitting of Mn states. The  Mn$^{3+}$ ion has a $3d^4_{\sigma}d^0_{\bar{\sigma}}$ electronic configuration with half-filled $t_{2g}$ and one electron in the doubly degenerate $e_g$ states, as illustrated schematically in \fref{fig:exchange_and_dos}~(i).  This electron resides on $d_{z^2}$ state rather than $d_{x^2-y^2}$ state to experience less Coulomb repulsion of O-$p$ electron. This leads to a  subsequent Jahn-Teller distortion which lifts the $e_g$ state degeneracy. Therefore, the Mn$^{3+}$ oxidation state and the ensuing Jahn-Teller distortion can be seen as a consequence of the electron-electron interaction on the Fe ions.

The Mn$^{3+}$ and Fe$^{3+}$ oxidation states and the Jahn-Teller distortion set the stage for AFM. 
If we assume the half-filled Fe-$d$ shell has down-spin, then it contributes in a superexchange interaction with the O-$p_{\uparrow}$ state since the Pauli principle prevents an extra down electron on Fe. The O-$p_{\downarrow}$ state, on the other hand,  contributes in a superexchange interaction with Mn inducing a spin-up configuration for Mn as shown in \fref{fig:exchange_and_dos} (i). Therefore,  magnetic moments on Mn and Fe align anti-parallel leading to a ferrimagnetic ground state for 
La$_2$MnFeO$_6$.

As mentioned above, the Jahn-Teller distortion and other structural distortions play a role in promoting the ferrimagnetic ground state. In A$_2$B'B''O$_6$ double perovskites, magnetic ions at B' and B'' sites are surrounded by oxygen ions in an octahedral environment. The oxygen octahedra can experience more or less tilting, depending on the ionic radius of the cations at A, B' and B'' sites. 
Octahedral tiltings lead to B'-O-B'' bonding angles that can differ from the ideal, cubic 180\degree~case.\cite{ref-a2bbo6} In the relaxed structures, the Mn-O-Fe bonding angles are $\simeq155$\degree~in the FM phase, and $\simeq153$\degree~in the AFM phase. In order to study the impact of structural distortion on the ground state, we generated structures with Mn-O-Fe bonding angles ranging from 150\degree~to 180\degree. We also studied structures with and without imposing Jahn-Teller distortion. 
Above $\simeq165$\degree, the ground state is ferromagnetic without  Jahn-Teller distortion. The oxidation states are Mn$^{4+}$ and Fe$^{2+}$. Below $\simeq165$\degree, as in the case of the relaxed structure, the ground state is antiferromagnetic with Jahn-Teller distortion. The oxidation states are Mn$^{3+}$ and Fe$^{3+}$. In the absence of Jahn-Teller distortion, the unoccupied Mn-$e_g$ state has a higher overlap with O-$p_{\uparrow}$ state, opening a hopping channel for this electron. In that case, O-$p_{\downarrow}$ can hop to Fe if the five d orbitals are occupied by up instead of down electrons. This spin configuration decreases the total energy of the FM configuration with spin-up Mn. This is confirmed by our ab initio calculations without Jahn-Teller distortion.

From the above discussion, one can see that ideal situation for a FM ground state in the $3d$-$3d$ double-perovskite oxides occurs when transition metals have $d^3$-$d^8$ configuration which happens in LMNO. This explains why LMNO has the largest Curie temperature, $T_c=280 K$, in this family. 

\section{Magnetic exchange couplings and Curie temperature}
\label{section:exchange}
Another important factor in designing magnetocaloric materials is the Curie temperature, $T_C$. Estimating $T_C$ from ab initio calculations is a challenging task. One approximate way, which we follow here, is to derive a spin Hamiltonian for the system and then evaluate its transition temperature. In the following subsection we present the method. The results follow.

\subsection{Extracting the exchange constants}
\begin{figure}
    \centering
    \begin{tabular}{ccc}
         \multicolumn{3}{c}{
         \includegraphics[width=0.56\columnwidth]{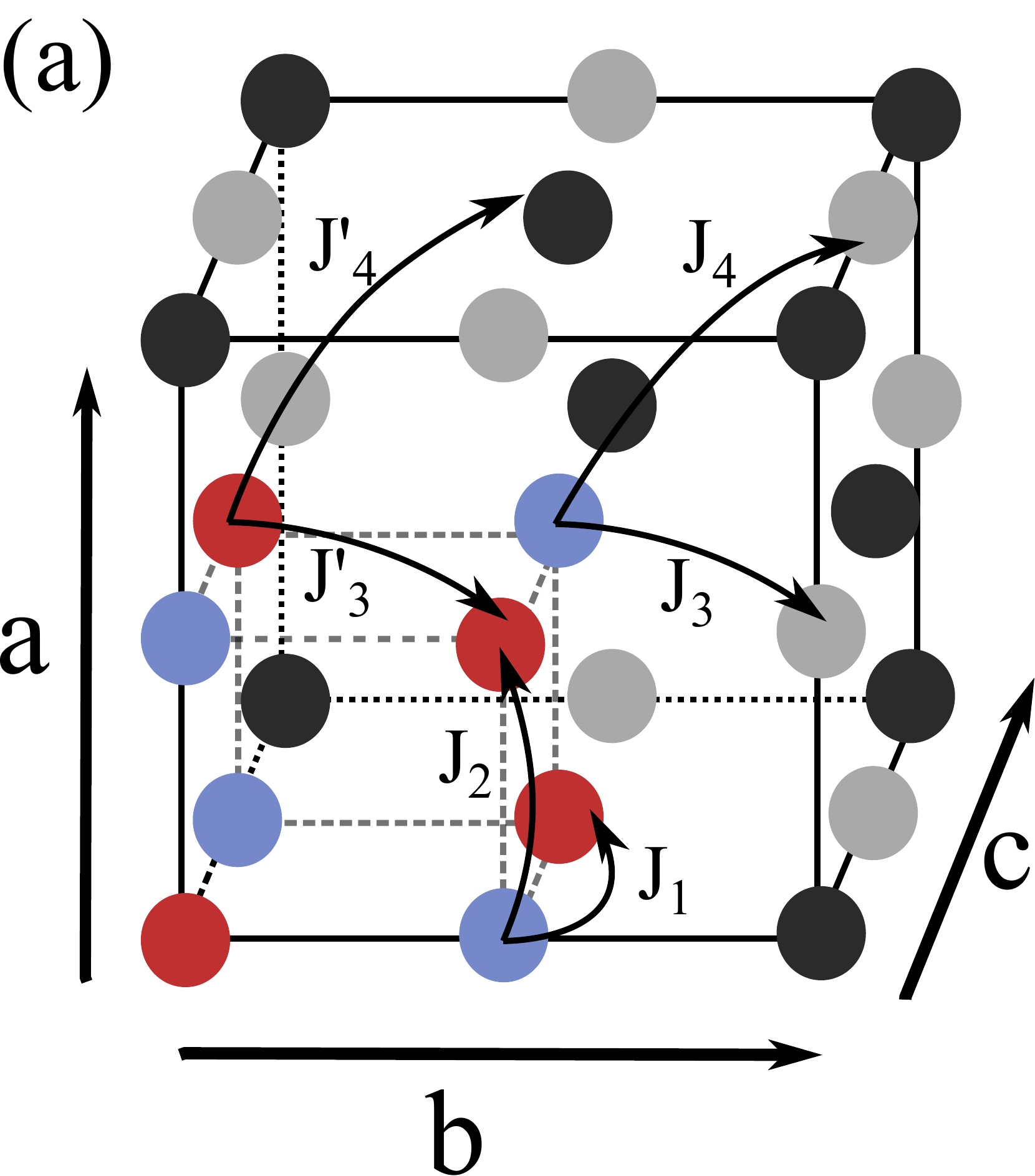}}\\
         \includegraphics[width=0.25\columnwidth]{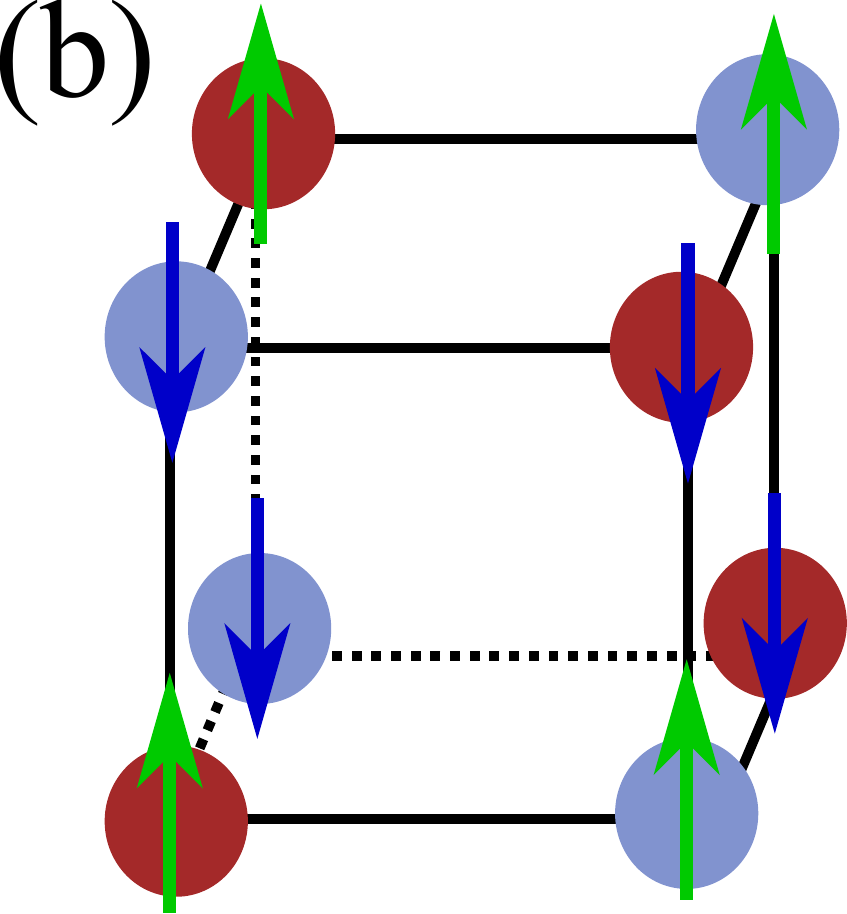} & \hspace{0.2cm} & 
         \includegraphics[width=0.25\columnwidth]{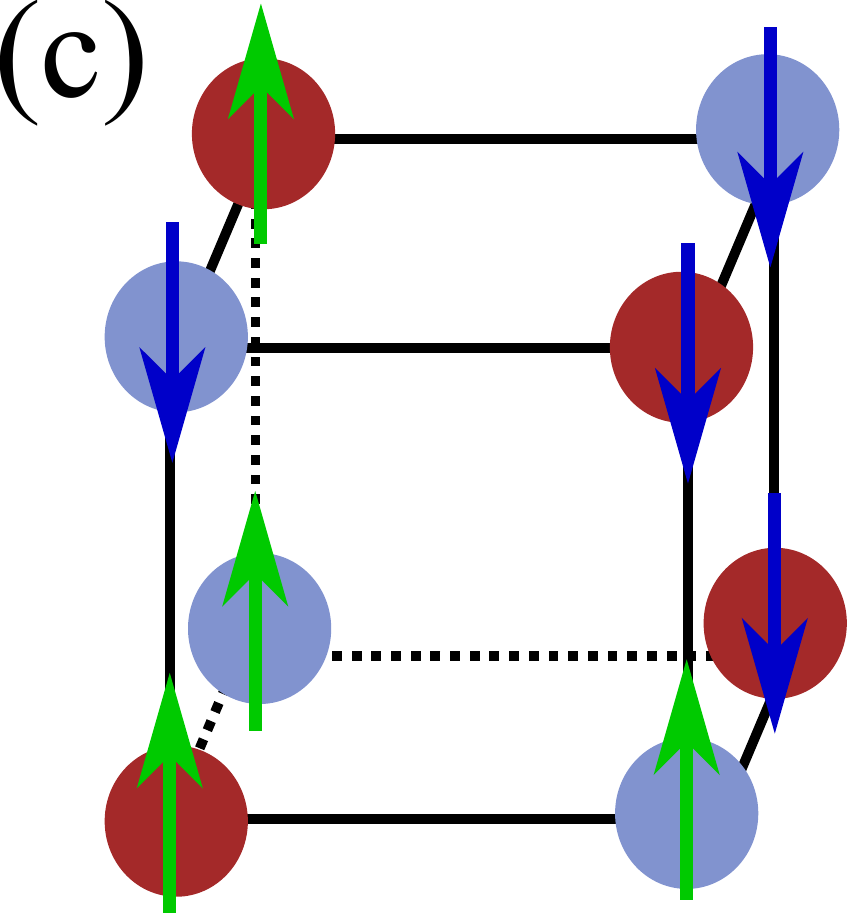} \\
         \includegraphics[width=0.25\columnwidth]{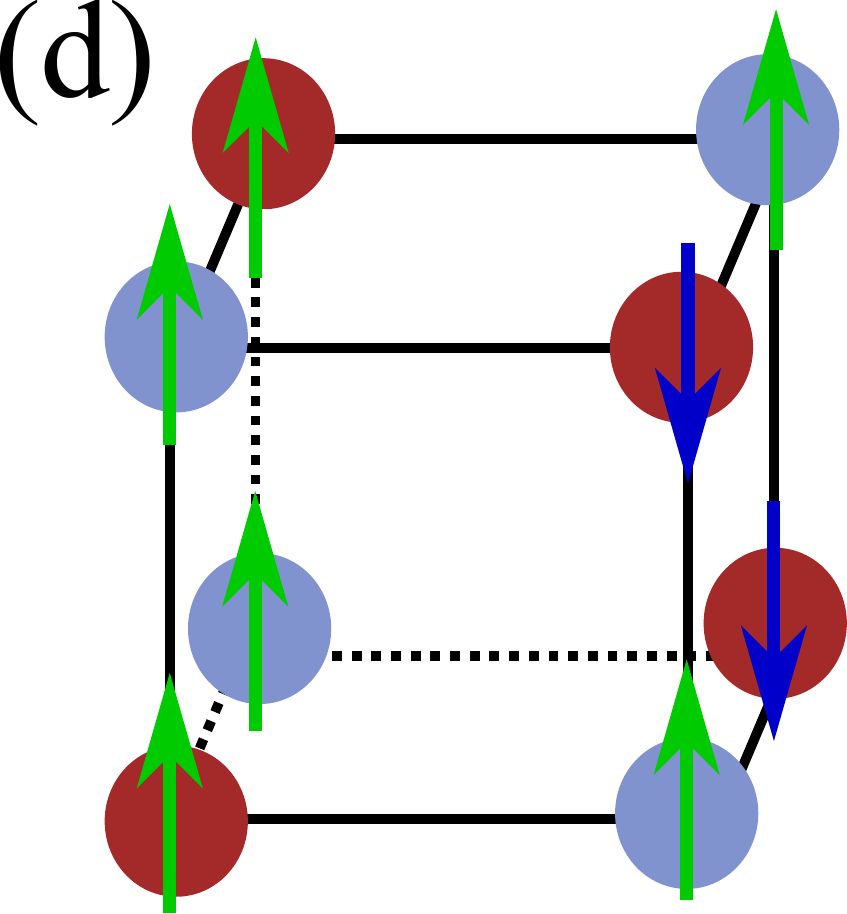} & &
    \includegraphics[width=0.25\columnwidth]{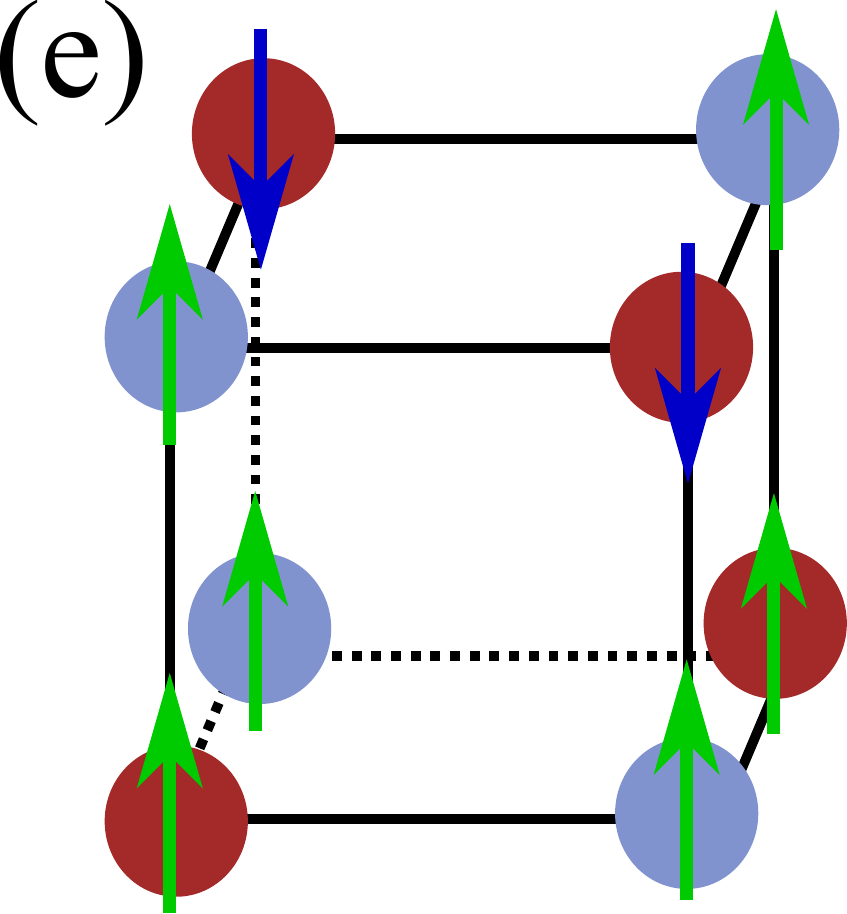} \\
    \end{tabular}
    \caption{(Color online) Top panel: (a) Reduced supercell with the $4$ non-equivalent Mn (blue) and Ni(Co, Fe) (red) atoms. The gray atoms are on the Mn sublattice and the black ones on the Ni(Co, Fe) sublattice. The lattice vector $\mathbf{a}$ denotes the out of plane direction, while the lattice vectors $\mathbf{b}$ and $\mathbf{c}$ generate the plane. Bottom panel: Spin configuration for Mn (blue) and Ni(Co, Fe) (red) sublattices in (b) AFM4, (c) AFM5, (d) AFM6 and (e) FiM phases.}
    \label{fig:uncommon_afm}
\end{figure}

We neglect the induced magnetic moment of oxygen because it is very small in comparison to  magnetic moments of the transition metal ions. The magnetic exchange interaction can be evaluated by mapping the DFT total energy to the Ising model,~\cite{PhysRevB.89.214414}
\begin{equation}
H=-\sum_{ij}J_{ij}S_i^zS_j^z, \label{eq:ising}
\end{equation}
where $J_{ij}$ is the exchange interaction between spins $S^z_i$ and $S^z_j$ residing at the $i$th and $j$th sites, respectively. Here, we assume that the exchange couplings have energy units, hence the spin values are divided by  the Bohr magneton. To obtain the long range exchange interactions, we first consider a $2\times 2\times 2$ supercell. The primitive unit cell contains two Mn ions and two   Ni(Co, Fe) ions, hence the supercell contains $32$ magnetic ions. We consider six independent exchange pathways connecting various Mn and Ni(Co, Fe) sites. $J_1$ and $J_2$ are the nearest-neighbor in-plane and out of plane couplings between Mn and Ni(Fe, Co), while $J_3$($J'_3$) and $J_4$($J'_4$) are the next nearest-neighbor in-plane and out of plane couplings between Mn(Ni, Fe, Co) magnetic moments. The exchange couplings are illustrated in \fref{fig:uncommon_afm} (a). 

In order to reduce the errors, we work with the total energy differences with respect to the ground state rather than the absolute total energy values. We fixed the structure and changed the magnetic order between eight different spin configurations and calculated their total energy. The spin configurations are chosen by adopting different configurations for each transition metal sublattice. Each sublattice can have the following spin configurations : (i) in-plane and out of plane FM (ii) in-plane FM and out of plane AFM (iii)  in-plane AFM and out of plane FM.  The chosen spin configurations are given in \tref{tab:spin_config}. Because of symmetries common to all the magnetic orders considered in the calculations, we were able to reduce the $2\times 2\times 2$ supercells to smaller ones that are twice as big as the primitive ones. These reduced supercells contain $4$ non-equivalent Mn and $4$ non-equivalent Ni(Fe, Co) atoms, hence a total of $8$ magnetic ions. 

We use GGA calculations to find the energy of each configuration in the LMNO and LMCO phases. In the case of LMFO, since GGA does not predict the right ground state, we use GGA+U with $U_{eff}=3.0$ eV in Mn-$3d$ and Fe-$3d$ shells to find the energies.
\begin{table}
\begin{center}
    \begin{tabular}{ l  c  c  c }
    \hline
    \hline
    Configuration & Sublattice B' & Sublattice B'' & Spin alignment \\ 
    & Mn & Ni, Co, Fe & \\ \hline
    FM & i & i & in phase\\ 
    AFM1 (G-type) & i & i & out of phase \\ 
    AFM2 (A-type) & ii & ii & in phase \\ 
    AFM3 (C-type) & ii & ii & out of phase \\ 
    AFM4 & iii & iii & (see caption)  \\ 
    AFM5 & ii & iii & n.a.  \\ 
    FiM & i & ii & n.a. \\ 
    AFM6 & i & iii & n.a.  \\ 
    \hline
    \hline
    \end{tabular}
    \caption{Spin configuration of the sublattices used in the 8 magnetic configurations. The transition metal sublattice spin configurations are : (i) in-plane and out of plane FM (ii) in-plane FM and out of plane AFM and (iii) in-plane AFM and out of plane FM. For AFM4, the spin alignment of the sublattices is chosen in such a way that the out of plane nearest neighbor alignment is AFM. Spin alignment of the sublattices in configurations AFM5, AFM6 and FiM does not influence the expression of the total energy since there is no net contribution of nearest neighbor (Mn-Fe/Co/Ni) interaction to the total energy. Uncommon configurations AFM4, AFM5, FiM and AFM6 are illustrated in \fref{fig:uncommon_afm} (b) to (e).}\label{tab:spin_config}
\end{center}
\end{table}

The corresponding total energy mapped to the Ising model can be obtained as follows. The primitive unit cell contains two Mn and two Ni (Co, Fe) ions. Each Mn (Ni, Co, Fe) has four in-plane and two out of plane nearest-neighbors which belong to the other sublattice. The number of in-plane and out of plane next-nearest neighbors are eight and four, respectively, and they belong to the same sublattice.  Therefore, letting $S^z$ and $S'^{z}$ denote the $z$-component of the spin on B' and B'' sublattices, the nearest-neighbor interaction contributes $-4J_1S^z S'^{z}-2J_2S^z S'^{z}$ per magnetic ion to the total energy  while the next-nearest-neighbor interaction contribution depends on the sublattice and is given by $-4J_3\mathcal{S}^z S^{z}-8J_4\mathcal{S}^z \mathcal{S}^{z}$ or $-4J'_3S'^{z} S'^{z}-8J'_4S'^{z} S'^{z}$ per magnetic ions.  Therefore, the total energies of the above spin configurations for primitive unit cell are:
\begin{align}
E_{tot}^{FM} = &-(16J_1+8J_2)\mathcal{S}^z\mathcal{S}'^z-(8J_3+16J_4)\mathcal{S}^{z2}\nonumber\\
&-(8J'_3+16J'_4)\mathcal{S}'^{z2},\\
E_{tot}^{AFM1} = &+(16J_1+8J_2)\mathcal{S}^z\mathcal{S}'^z-(8J_3+16J_4)\mathcal{S}^{z2}\nonumber\\
&-(8J'_3+16J'_4)\mathcal{S}'^{z2},\\
E_{tot}^{AFM2} = &-(16J_1-8J_2)\mathcal{S}^z\mathcal{S}'^z-(8J_3-16J_4)\mathcal{S}^{z2}\nonumber\\
&-(8J'_3-16J'_4)\mathcal{S}'^{z2},\\
E_{tot}^{AFM3} = &+(16J_1-8J_2)\mathcal{S}^z\mathcal{S}'^z-(8J_3-16J_4)\mathcal{S}^{z2}\nonumber\\
&-(8J'_3-16J'_4)\mathcal{S}'^{z2},\\
E_{tot}^{AFM4} = &+8J_2\mathcal{S}^z\mathcal{S}'^z+8J_3S^{z2}+8J'_3\mathcal{S}'^{z2},\\
E_{tot}^{AFM5} = &-(8J_3-16J_4)\mathcal{S}^{z2}+8J'_3\mathcal{S}'^{z2},\\
E_{tot}^{FiM} = &-(8J_3+16J_4)\mathcal{S}^{z2}-(8J'_3-16J'_4)\mathcal{S}'^{z2},\\
E_{tot}^{AFM6} = &-(8J_3+16J_4)\mathcal{S}^{z2}+8J'_3\mathcal{S}'^{z2},
\label{eq:energies}
\end{align}
where $\mathcal{S}^z$ and $\mathcal{S}'^z$ denote the extreme values of the spins.
These equations should be multiplied by $2$ to obtain the total energy of the supercell considered here.  
They yield seven energy differences; six of them are used to calculate the exchange couplings and the last one is used to verify their validity.

\subsection{Calculated exchange constants and mean-field Curie temperatures}
The calculated magnetic exchange interactions for LMNO are presented in \tref{tab:js}. The nearest-neighbor exchange couplings are FM-type and sizable while the next-nearest-neighbor magnetic moments are coupled antiferromagnetically. The rather short-range magnetic interaction is a consequence of the rather localized $3d$ electron wave functions.  The next-nearest neighbors are identical ions with half-filled shells and their AFM coupling can be understood in terms of a simple Hubbard model: The AFM alignment of the magnetic moment allows for hopping of the electron between these sites and reduces kinetic energy. 
%
%
\begin{table}
\begin{center}
    \begin{tabular}{c  c c c  c }
    \hline
    \hline
     & Interaction path&\multicolumn{3}{c}{Values (meV)}\\ 
     &  & LMNO &LMCO&LMFO \\ \hline
         $J_1$ & Mn-B'' (in plane) &  4.51 & 3.36 & -1.75 \\
         $J_2$ & Mn-B'' (out of plane) &   4.09 & 2.85  & -0.97 \\
         $J_3$ & Mn-Mn (in plane) &   -0.15 & -0.59  & -0.04 \\
         $J_4$ & Mn-Mn (out of plane) &  -0.10  & -0.45  & -0.28   \\
         $J'_3$ & B''-B'' (in plane)&   -0.04   &  0.28  &  0.16 \\
         $J'_4$ & B''-B'' (out of plane) &  -0.11  & 0.08  & -0.18 \\
    \hline \hline
    \end{tabular}
    \caption{ Calculated magnetic exchange interactions for LMNO, LMCO and LMFO. Positive (negative) value denotes FM (AFM) coupling.  The extreme values of the spins are $\mathcal{S}^z_{\rm Mn} = 3/2$,  $\mathcal{S}'^z_{\rm Ni} =1$, $\mathcal{S}'^z_{\rm Co} =3/2$ in LMNO and LMCO, and $\mathcal{S}^z_{\rm Mn} = 2$ and $\mathcal{S}'^z_{\rm Fe} =5/2$ in LMFO. 
    }\label{tab:js}
\end{center}
\end{table}
%
%

The exchange couplings for LMCO are also listed in \tref{tab:js}. One can notice that the Mn-Co couplings are smaller than the Mn-Ni ones, which is in agreement with the discussion of the superexchange mechanism in these materials of \sref{section:LMNO_LMCO} and \sref{section:LMFO}. We argued that, due to the superexchange interaction, the ideal situation to promote an FM phase over an AFM one in the family of compounds studied is the $3d^3$-$3d^8$ case. This situation occurs in LMNO and it could explain why the exchange couplings are larger in this material, and also why its experimental $T_C$ is the largest in this family even though its total moment is not the largest. Once again, the next-nearest neighbor couplings are smaller than $J_1$ and $J_2$ by one or two orders of magnitude due to the localization of $3d$ wave functions. 
Note that Co-Co exchange couplings are FM. In contrast with LMNO, in which direct hopping between Ni ions only occurs between $e_g$ states, in LMCO the direct hopping processes between $t_{2g}$ states are allowed as well. Hence, a simple argument based on half-filled Hubbard model is not applicable to Co ions.

In the case of LMFO, Mn and Fe magnetic moments are coupled antiferromagnetically, with $|J_1|$ and $|J_2|$ smaller than what we found previously for LMNO and LMCO. Finally, the Curie temperatures evaluated from a mean-field treatment of the Ising model are given in \tref{tab:temperatures}. For the details of the mean-field calculation, see appendix~\ref{app:MF}.  The mean-field theory overestimates the Curie temperatures. However, it gives the correct trend. 

Finally, in order to examine the accuracy of the exchange couplings, we used them to evaluate the energy differences of a new magnetic configuration with respect to FM state and compare the results with those found directly from the ab initio calculations.  The configuration AFM6,   given in \tref{tab:spin_config}, was not used in the calculation of the couplings. The energy difference from the ab initio calculations and from the Ising model using exchange couplings from \tref{tab:js} are listed in \tref{tab:energy_afm6}. The agreement is very good in the case of LMNO (less than $1$\% of disparity), and relatively good in the cases of LMCO and LMFO (respectively $5$\% and $11$\% of disparity).

\begin{table}
    \centering
    \begin{tabular}{c c c c}
    \hline
    \hline
         & \multicolumn{3}{c}{$E_{AFM6}-E_{FM}$} \\
         &  From $J$ values & & From ab initio calculations \\
         \hline
         LMNO & 0.154 & & 0.155 \\
         LMCO & 0.176 & & 0.185 \\
         LMFO & -0.181 & & -0.202 \\
         \hline
         \hline
    \end{tabular}
    \caption{Energy difference between AFM6 phase and FM phase in eV.}
    \label{tab:energy_afm6}
\end{table}
\section{Transition temperature from Monte-Carlo calculations}
\label{section:CMC}

The mean field treatment of the spin Hamiltonian does not take into account fluctuations hence overestimates the Curie temperatures $T_C$. In order to get better approximations, we perform classical Monte-Carlo calculations with the GT-GPU method on a cubic lattice\cite{2018arXiv180109379B} using the $J$ and extreme spin values listed in \tref{tab:js}.  

\subsection{Monte Carlo methodology}
Genetic Tempering  is a Monte-Carlo algorithm that uses many copies of the Metropolis-Hastings algorithm \cite{metropolis1953equation,hastings1970monte,landau2014guide} to generate highly precise statistical measurements and eliminate auto-correlation error.  To ensure accuracy, initial state samples  are chosen from unconverged samples  surrounding the target answer.  Running another round of Monte-Carlo sampling on these initial states  gives a partial cancellation of the relaxation error.~\footnote{ Here $10^5$ uncounted steps and $5\times10^3$ counted steps were used in initial state samples, while $5\times 10^5$ uncounted steps with $10^6$ counted steps were used in calculating the Binder's cumulant. Random numbers were made with the counter-based random number generator of Ref.~\onlinecite{Random123}. All data for a single system size were obtained with one graphics processing unit and one computing processing unit in $1-2$ hours or less. }
The $T_c$ is obtained from Binder's cumulant, \cite{landau2014guide}
$B(T,L)=1-\langle M^4\rangle_L/(3\langle M^2\rangle_L^2)$,
with $L$ the linear lattice size. In the thermodynamic limit, $L \rightarrow \infty$, the Binder's cumulant tends towards $2/3$ for $T<T_c$ and 0 for $T>T_c$, hence it is discontinuous in this limit. In simulation with finite $L$, the intersection point of the Binder's cumulants for different system sizes determines $T_c$ and usually depend only rather weakly on those sizes. 

\subsection{Improved Curie temperatures}
The Curie temperatures obtained from Monte-Carlo are shown in \tref{tab:temperatures} and compared with mean-field values and experimental ones.  The transition temperature of bulk ordered LMFO is not known but the magnetization measurements indicate a lower $T_C$ than LMCO, as seen in \fref{fig:MvsT}.  As can be seen from \tref{tab:temperatures},  both mean-field and Monte-Carlo give consistent trend with experimental data but overestimate them. The Monte-Carlo prediction is closer to experimental transition temperatures. It is worth mentioning that Curie temperature of LMNO and LMCO are affected by the (AF) next-nearest neighbor exchange  coupling: they have smaller magnitude in comparison  with nearest neighbor  exchange coupling, but the number of next-nearest neighbor is larger. Finally, capturing the correct trend allows both mean-field and Monte-Carlo methods to be used reliably  in material design.

\begin{table}
    \centering
    \begin{tabular}{cccc}
    \hline
    \hline
          &  \multicolumn{3}{c}{Transition temperature (K)}\\
         & Mean-field & Monte-Carlo & Experimental \\
         \hline
         LMNO & 527 & 419 & 280 [\onlinecite{ADMA:ADMA200500737}] \\
         LMCO & 505 & 399 & 226 [\onlinecite{PhysRevB.67.014401}] \\
         LMFO & 418 & 329 & \\
         \hline
         \hline
    \end{tabular}
    \caption{Magnetic phase transition temperature in Kelvin obtained from different methods for LMB''O.}
    \label{tab:temperatures}
\end{table}

\section{Concluding remarks}
\label{Sec:Conclusion}
We used GGA and GGA+U calculations in order to understand the ground state electronic and magnetic properties of double perovskite La$_2$MnFeO$_6$. This material is predicted to be an insulating ferrimagnet, unlike similar compounds La$_2$MnCoO$_6$ and La$_2$MnNiO$_6$, which are insulating ferromagnets. The present study helped us understand the important role played by electron-electron interactions in the determination of the ground state of this material. We also showed that the interplay between crystal field, Mn-O-Fe bonding angle and electron-electron interactions also needs to be taken into consideration when analyzing the superexchange mechanism in LMFO. 

Large electron-electron interactions in Fe $3d$ shells promote Mn$^{3+}$ and Fe$^{3+}$ oxidation states and AFM ground state in LMFO. In contrast, we also saw that Mn-O-Fe bonding angles superior to $\simeq165$\degree~promote a FM ground state with Mn$^{4+}$ and Fe$^{2+}$ oxidation states. From these observations, we believe that studying two different new materials could be of interest in the global topic of magnetic refrigeration. 

First, in order to avoid the strong electronic repulsion in $3d$ shells that promote an AFM ground state in LMFO over an FM one, one could study the double perovskite La$_2$MnRuO$_6$ (LMRO). Valence orbitals in Ru are $4d$, which are more extended in space than Fe-$3d$ orbitals. Smaller correlations in LMRO could lead to a FM ground state with Mn$^{4+}$ and Ru$^{2+}$ oxidation states. However, experiments on disordered LMRO (space group $Pnma$) have found this material to be a ferrimagnet with trivalent Mn in high-spin configuration and trivalent Ru in low-spin configuration.~\cite{KAMEGASHIRA2000L6} The effect of the ordering of Mn and Ru atoms still needs to be investigated. 

The second option is to get rid of an excess electron in Fe by hole-doping the compound, which could lead to Mn$^{4+}$ and Fe$^{3+}$ oxidation states. If both Mn and Fe are in high-spin configuration, these oxidation states would lead to a ferromagnetic superexchange interaction. Moreover, the substitution of La$^{3+}$ cations with larger divalent cations would increase the tolerance factor of the material, which would in turn reduce the amount of octahedral tilting. This avenue will be explored further in work to come.

\begin{acknowledgments}
We are indebted to Michel C\^ot\'e for numerous discussions. This work has been supported by the Natural Sciences and Engineering Research Council of Canada (NSERC) under grant RGPIN-2014-04584, by the Canada First Research Excellence Fund, and by the Research Chair in the Theory of Quantum Materials. T.E.B. graciously thanks funding provided by the postdoctoral fellowship from Institut quantique. T.E.B. thanks Yehua Liu, Glen Evenbly, and David Poulin for discussions. 
Ab initio simulations and Monte Carlo calculations were performed on computers provided by the Canadian Foundation for Innovation, the Minist\`ere de l'\'Education des Loisirs et du Sport (Qu\'ebec), Calcul Qu\'ebec, Compute Canada and the Fonds de recherche du Qu\'ebec - Nature et technologies (FRQ-NT). 
\end{acknowledgments}

\appendix
%
\section{Mean-field theory for Curie temperature}\label{app:MF}
In this section we recall mean-field theory for double perovskites, which allows us to estimate the Curie temperature. The magnetic system of LMNO, LMCO and LMFO can be modeled by the Ising model, \eref{eq:ising}.  The mean field approximation is to replace the configurational energy by the energy of a non-interacting system of spins each experiencing a field $h_{MF}$. The mean field Hamiltonian can be obtained by substituting $S^z_i = \langle S^z_i \rangle + \delta S^z_i$ with $\delta S^z_i \equiv S^z_i - \langle S^z_i \rangle$ and neglecting terms of order $(\delta S^z_i )^2$ in  \eref{eq:ising}. For the FM configuration, the mean-field energy is
\begin{align}
E^{FM}_{MF} &= - \sum_i h_{MF} S^z_i - \sum_j h'_{MF} S'^z_j,\\
h_{MF} &= 2(4J_1+2J_2)m'+2(4J_3+8J_4)m,\\
h'_{MF} &= 2(4J_1+2J_2)m+2(4J'_3+8J'_4)m',
\end{align}
where magnetizations per spin are $m=(1/2N)\sum_i \langle S^z_i \rangle$ and $m'=(1/2N)\sum_i \langle S'^z_i \rangle$ and $2N$ denotes the total number of sites of a given type. Recall that $h_{MF}$ and $h'_{MF}$ depend on temperature through the temperature dependence of  $m$ and $m'$.

The magnetization per spin is given by $m=\sum_{S^z_i}p(S^z_i)S^z_i$ where $p(S^z_i)$ denotes the single-spin Boltzmann distribution 
\begin{align}
p(S^z_i) = (\frac{1}{\mathcal{Z}})\exp(\beta h_{MF}S^z_i),
\end{align}
and $\mathcal{Z}$ is the partition function
\begin{align}
\mathcal{Z} &= Z\cdot Z'\nonumber \\
&=\sum_{S^z_i} \exp(\beta h_{MF}S^z_i)\cdot\sum_{S'^z_i} \exp(\beta h'_{MF}S'^z_i). 
\end{align}

It is straightforward to show that the magnetizations are given by Brillouin functions
\begin{align}
m(T)&=\frac{1}{\beta}\frac{\partial \ln (Z)}{\partial h_{MF}}=-\frac{1}{2}\coth(\beta h_{MF}/2) \nonumber\\
&+(\mathcal{S}^z+\frac{1}{2})\coth(\beta h_{MF}(\mathcal{S}^z+1/2)),\\
m'(T)&=\frac{1}{\beta}\frac{\partial \ln (Z')}{\partial h'_{MF}}=-\frac{1}{2}\coth(\beta h'_{MF}/2) \nonumber\\
&+(\mathcal{S}'^z+\frac{1}{2})\coth(\beta h'_{MF}(\mathcal{S}'^z+1/2)),
\end{align}
where $\mathcal{S}^z$ and $\mathcal{S}'^z$ denote the extreme values of the spins. For example, for LMNO, they are $\mathcal{S}^z_{\rm Mn} = 3/2$ and $\mathcal{S}'^z_{\rm Ni} =1$. Note that one could have used $\mathcal{S}^z_{\rm Mn} = 3$ and $\mathcal{S}'^z_{\rm Ni} =2$ which renormalizes the exchange couplings but leaves Curie temperature invariant. 
These coupled equations can be solved to obtain $m$ and $m'$. For high $T$ (low $\beta$), the only solution is $m=m'=0$ whereas for low $T$ there are possible non-trivial solutions. The solution with non-zero $|m|$ and $|m'|$ appears at Curie temperature. By expanding the $\coth$ functions in the limit of small $h_{MF}$ and $h'_{MF}$, the above equations give the eigenvalue problem
\begin{align}
m(T)&=\frac{\mathcal{S}^z(\mathcal{S}^z+1)}{3k_BT} h_{MF}(T),\\
m'(T)&=\frac{\mathcal{S}'^z(\mathcal{S}'^z+1)}{3k_BT} h'_{MF}(T),
\end{align}
which yields the following equation for Curie temperature:
\begin{widetext}
\begin{align}
3k_BT_c = [\mathcal{S}^z(\mathcal{S}^z+1)h_2+\mathcal{S}'^z(\mathcal{S}'^z+1)h'_2] +
\sqrt{4h_1^2\mathcal{S}^z(\mathcal{S}^z+1)\mathcal{S}'^z(\mathcal{S}'^z+1)+[\mathcal{S}^z(\mathcal{S}^z+1)h_2-\mathcal{S}'^z(\mathcal{S}'^z+1)h'_2]^2},
\label{eq:Tc}
\end{align}
\end{widetext}
where we have defined $h_1= 4J_1+2J_2$, $h_2= 4J_3+8J_4$ and $h'_2= 4J'_3+8J'_4$. Note that in the case with identical moments, $\mathcal{S}^z=\mathcal{S}'^z$, and only nearest neighbor coupling, the above equation reduces to textbook formula, i.e., $3k_BT_c = 2h_1 \mathcal{S}^z(\mathcal{S}^z+1)$.

Assuming that LMFO undergoes a ferrimagnetic to paramagnetic transition at N\'eel temperature $T_N$, then $T_N$ can also be obtained from \eqref{eq:Tc}. In this case, the only difference between the treatment of $T_N$ and $T_C$ is the sign of nearest-neighbor interaction $h_1$, which also appears in $h_{MF}$ and $h'_{MF}$. However, the sign of $h_1$ disappears in \eqref{eq:Tc}.  For the classical Ising model on bipartite lattices, it is well known that a simple relabeling of up and down for one of the sublattices maps the nearest-neighbor ferromagnetic model to the antiferromagnetic one .
%

\end{document}